\documentclass[12pt,draftclsnofoot,onecolumn]{IEEEtran}
\usepackage{amssymb}
\usepackage{amsfonts}
\usepackage{epstopdf}
\usepackage{amsfonts,subfigure,multicol,color,verbatim,
graphicx,cite,epsfig,amssymb,amsmath,cases,bm,algorithm,
algorithmic,xcolor,multirow,array}

\begin{document}

\title{Economical Energy Efficiency ($E^3$): An Advanced Performance Metric for 5G Systems}
\author{Zhipeng~Yan, Mugen~Peng, \IEEEmembership{Senior Member,~IEEE}, and Chonggang~Wang, \IEEEmembership{Senior Member,~IEEE}
\thanks{Zhipeng~Yan (e-mail: {\tt yanzhipeng1981@126.com}) and Mugen~Peng (e-mail: {\tt pmg@bupt.edu.cn}) are with the Key Laboratory of Universal Wireless Communications for Ministry of Education, Beijing University of Posts and Telecommunications, Beijing, China.
Chonggang~Wang (e-mail: cgwang@ieee.org)
is with the InterDigital Communications, King of Prussia, PA, USA.}
\thanks{This paper was supported in part by the National High
Technology Research and Development Program (863 Program) of China under Grant No. 2014AA01A707, the National Program for Support of Top-notch Young Professionals, the National Natural Science Foundation of China (Grant No.61271198), and the Science and Technology Development Project of Beijing Municipal Education Commission of China (Grant No. KZ201511232036).}

\thanks{Copyright (c) 2016 IEEE. Personal use of this material is permitted, but republication/redistribution requires IEEE permission.
See http://www.ieee.org/publications standards/publications/rights/index.html for more information. }
}

\maketitle

\begin{abstract}
The performances of the fifth generation (5G) wireless communication systems are significantly affected by edge cache and transport network. These emerging components bring substantial costs of the placement and utilization, and the evaluation of the cost impact is beyond the capability of traditional performance metrics, including spectral efficiency (SE) and energy efficiency (EE). In this article, economical energy efficiency ($E^3$) is proposed, whose core idea is to take SE/EE and cost into account to evaluate comprehensive gains when different kinds of advanced technologies are used in 5G systems. The $E^3$ results are shown when the transport network and edge cache are separately or jointly used. Open issues in terms of modeling the cost, $E^3$ optimization based radio resource allocation, and $E^3$ optimization for internet of things, are identified as well.
\end{abstract}


\newpage

\section{Introduction}

The rapid development of mobile internet and internet of things (IoT) is now on a path to exhaust the capabilities of existing wireless systems, which motivates a growing number of researchers to develop the fifth generation (5G) wireless systems. Besides capacity and greenness, high values are also put on some other aspects in this process, including scalability, flexibility, and interoperability, which brings challenging requirements on the capability of 5G systems. Given all these pluralistic requirements, it has now become a consensus that compared with the traditional wireless communication systems, 5G systems need to be capable of supporting 10 to 100 times higher typical user data rate, connecting 10 to 100 times higher number of devices, and ensuring 5 times reduced end-to-end latency \cite{Definition_of_5G}.

\begin{figure}[t]
\begin{center}
\vspace*{0pt}
\includegraphics[scale=0.55]{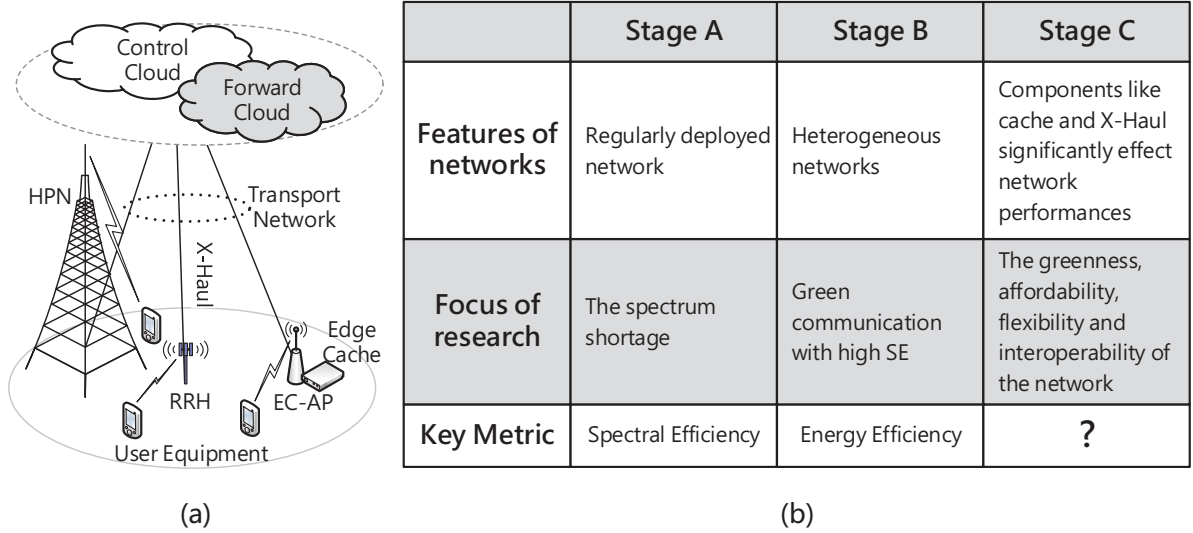}
\caption{Challenges for performance metrics. (a) An example of 5G systems. (b) Evolution of performance metrics for mobile networks.}
\label{EvolutionOfNetwork}
\end{center}
\end{figure}

To handle such challenges, several innovative network architectures have been proposed as prospective solutions for 5G systems, such as cloud radio access networks (C-RANs) \cite{Definition_of_CRAN}, heterogeneous cloud radio access networks (H-CRANs) \cite{Definition_of_HCRAN}, and fog computing based radio access networks (F-RANs) \cite{Definition_of_FRAN}. The performances of all these potential paradigms strictly depend on the capability of transport networks, which are composed of backhaul and fronthaul (X-Haul). Besides, edge cache is widely used in F-RANs to alleviate the burden on transport networks and reduce end-to-end latency. Fig. \ref{EvolutionOfNetwork} (a) shows an example of F-RAN based 5G systems consisting of high power nodes (HPNs), remote radio heads (RRHs), and edge-computing access points (EC-APs). Given the importance of X-Haul and edge cache, it is indispensable to fully consider the corresponding impacts of these emerging components.

While making comparisons among different X-haul and edge cache strategies, or setting explicit long term targets for the development of 5G systems, appropriate metrics that enhancing traditional spectral efficiency (SE) and energy efficiency (EE) are urgent since they provide the quantified evaluation of the results. As Stage A shown in Fig. \ref{EvolutionOfNetwork} (b), SE, defined as the system throughput per unit of bandwidth, is widely used for the traditional cellular networks with regular deployment, in which the key issue is the shortage of spectrum resources. As the techniques keep evolving, much attention is paid to the greenness issues. Taking no consideration on energy consumptions, SE fails to characterize the energy-saving feature of heterogeneous networks (HetNets). Subsequently, EE, defined as the system throughput per unit of energy consumption, is proposed as a solution. Since it evaluates network performance with both throughput and energy consumption taken into consideration, EE well fits the demands of green communication, and is widely used in green networks, as Stage B shown in Fig. \ref{EvolutionOfNetwork} (b).

The performance requirements of 5G systems become increasingly pluralistic, and they need to be evaluated in a comprehensive manner. Since the affordability is regarded as an essential concern of 5G systems, much effort should be focused on reducing the total cost, including at least the overheads on network infrastructure, deployment, operation, and management due to using X-Haul and edge cache \cite{WhitePaper_5G_Initiative}. With the additional cost of 5G systems, the traditional SE and EE metrics appear to be insufficient to evaluate the impacts by this kind of cost-related techniques. As a result, as Stage C shown in Fig. \ref{EvolutionOfNetwork} (b), it is urgent to propose an advanced performance metric to comprehensively evaluate both EE and the cost induced by various advanced components of 5G systems.

Accurately, there have already been some research works focusing on the cost issue due to introducing the X-Haul and edge cache. A brief analysis of the relationship between SE and EE is discussed in \cite{DE_Optimization_That_Take_EE_Into_Consideration}, with the conclusion that the relation is not a simple trade-off, and the relation under realistic and complex network scenarios still requires further investigation. In 5G systems, this kind of trade-off should be revisited and the impact of cost should be highlighted. In \cite{Benefit_of_Wireless_Backhaul}, the way to reduce the cost on X-Haul is proposed, and the authors of \cite{Cache_HitRatio} analyze cost-effective caching strategies. In \cite{DE_Optimization_Thesis}, the cost effective deployment strategies for HetNets are presented, where the cost efficiency is proposed as the ratio of system throughput to the total cost of the system. A study on the constitution of the cost is given in \cite{DE_EE_Relation_01_Balance}, together with deployment strategies to separately minimize the overall system cost and the total energy consumption, in which the cost and the greenness issues are not jointly considered.

To characterize scalability, flexibility, and interoperability in 5G systems, this article provides a feasible performance metric to show the comprehensive gains. In particular, an advanced performance metric named economical energy efficiency ($E^3$) is proposed, with the core idea of taking EE and cost into account. Based on the proposed metric, the impacts of X-Haul and edge cache are characterized. Numerical results imply that there are potential ways to optimize the $E^3$ performance of 5G systems, including proper X-Haul deployment strategies, advanced edge caching strategies, and the joint optimization of X-Haul and edge caching strategies. Challenges and open issues related to $E^3$ metric, including the cost model, $E^3$ optimization based radio resource allocation, and $E^3$ optimization for IoTs, are discussed as well.

The rest of this paper is organized as follows. The definition of the proposed $E^3$ metric is introduced in Section \uppercase\expandafter{\romannumeral2}, together with its characteristics. In Section \uppercase\expandafter{\romannumeral3}, after briefly exploiting the impact of X-Haul and edge cache, several numerical results of using the proposed metric are presented. Open issues are then discussed in Section \uppercase\expandafter{\romannumeral4}, followed by the Conclusion Section.

\section{Definition and Characteristics of $E^3$ Metric}

Instead of traditional ways with high complexity working on the relations between SE and EE \cite{DE_Optimization_That_Take_EE_Into_Consideration}, the proposed metric provides a feasible method to jointly consider the impacts on SE, EE, and cost. The core idea of the proposed $E^3$ metric is to take EE and cost into account to show comprehensive gains when different kinds of advanced techniques are used in 5G systems.

\subsection{Definition of $E^3$ Metric}

$E^3$ is defined as the ratio of effective system throughput to energy consumption weighted by cost coefficient, i.e., $ E^{3}=\frac{\sum\limits_{k \in \mathcal{K}}\alpha_k \, R_k}{\sum\limits_{n \in \mathcal{N}} (P_{Tn} + P_{0n} \, C_n)}$, in which $\mathcal{N}=\{1,2,3,...,N\}$ represents the set of all base stations (BSs) (including RRHs, EC-APs, and etc.), and $\mathcal{K}=\{1,2,3,...,K\}$ represents the set of all active user equipment (UEs). With $R_k$ representing the effective throughput of UE \emph{k}, and $\alpha_k$ representing the weight factor which is related to the priority, the effective system throughput is calculated as the sum of $R_k \cdot \alpha_k$ for all $k \in \mathcal{K}$. With $P_{Tn}$ representing the dynamical power dependant to the load of BS $n$, $P_{0n}$ representing the static power of BS $n$, and $C_n$ representing the cost coefficient, the weighted energy consumption is calculated as the sum of $P_{Tn} + P_{0n} \, C_n$ for all $n \in \mathcal{N}$.

Let $S(n)$ denote the set of BSs of the same kind as BS $n$, and these BSs are supposed to have similar edge cache resources and X-Haul solutions. With $\tilde{C}_{S(n)}$ representing the average cost for BSs belong to $S(n)$, and $C_0$ the benchmark cost, which can be calculated according to a certain referential BS (e.g., the most costly kind of BS in the system), $C_n$ is calculated as the ratio of $\tilde{C}_{S(n)}$ to $C_0$. To be clear, both $\tilde{C}_{S(n)}$ and $C_0$ are calculated according to unit covering area. Since $C_n$ is dimensionless, the unit of $E^3$ is ``bit/Joule", and $E^3$ can be regarded as an enhanced EE.

Note that though there have already been mature methods to calculate the throughput and the energy consumption, the way to calculate the cost is still not straightforward. The overall cost of 5G systems is determined by a large number of components, such as the infrastructure purchase, site installation, site operation, network optimization/maintenance, edge cache placement, X-Haul configuration, and content deliver to the edge cache. Other overheads due to the centralized/distributed cloudization also have a significant impact on the cost efficient. In addition, if the component of 5G systems is inherited from the existing cellular systems, the corresponding cost should be calculated differently.

\subsection{The Characteristics of $E^3$ Metric}

According to the definition, with normalized $C_n$, $E^3$ is equivalent to traditional EE metric, which implies that differences in cost are ignored with EE. For the proposed $E^3$ metric, when X-Haul and edge-caching techniques are used in 5G systems, the throughput is jointly considered with requirements on energy and budget. Although it affects neither throughput nor energy consumption, the inheritance of components from existing networks is beneficial to the $E^3$ optimization, as it reduces the cost.

\begin{figure}[t]
\centering
\vspace*{0pt}
\includegraphics[scale=0.65]{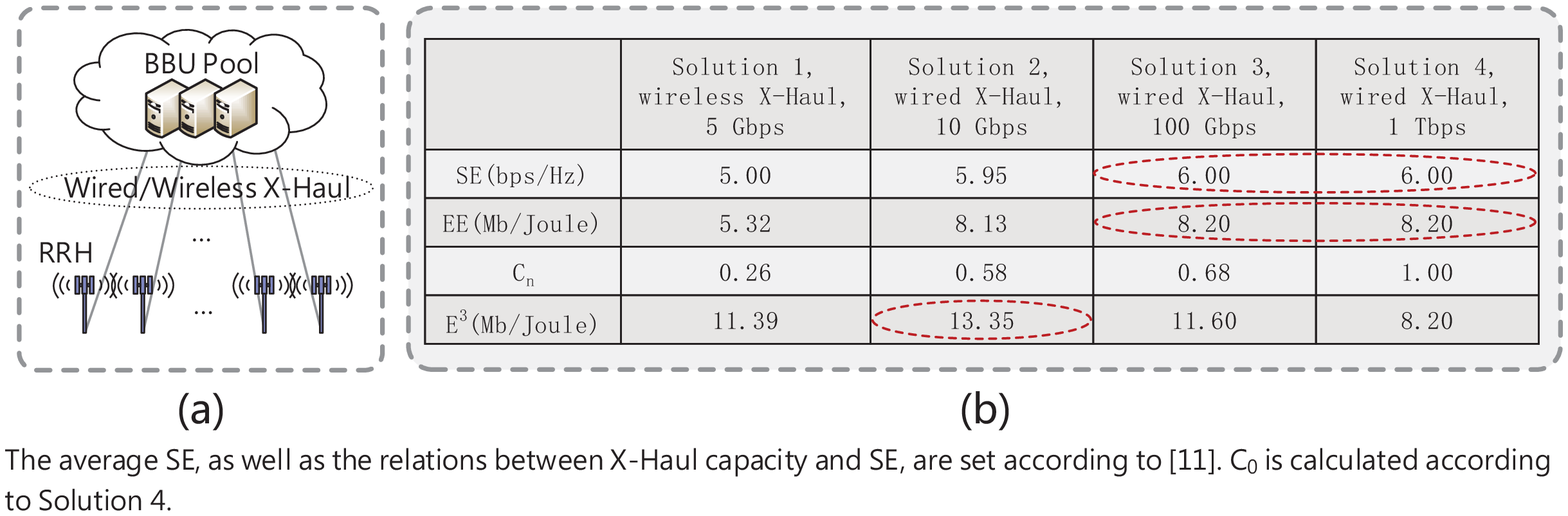}
\caption{Different performance metric comparisons: the impact of X-Haul. (a) An example of C-RAN based 5G systems. (b) Numerical results for different X-Haul solutions.}
\label{Xhaul}
\end{figure}

As an example, Fig. \ref{Xhaul} shows a comparison among different performance metrics. Without further notes, the SE-related numerical results in this article are obtained under the same configuration settings as that used in \cite{Reference_Numerical_Results_For_CRAN}. The energy-related parameters of the low power nodes (LPNs, including RRHs and EC-APs) are set according to the pico BS with 10 MHz bandwidth presented in \cite{EARTH_Report}. To be clear, $P_{Tn}$ for an AP with wireless X-Haul is assumed to be composed of two parts: the transceiver part and the wireless X-Haul part, in which the latter is assumed to be 3 times bigger than the preceding \cite{Reference_Numerical_Results_For_CRAN}. The parameter $C_n$ ranges from 0.26 to 1, depending on the used X-Haul solution. Those marked by dash lines are the best performances in the comparisons. According to the numerical results, for the given X-Haul solutions with different capacity and cost, $E^3$ can present the harm of over pursuit of large capacity X-Haul, while both SE and EE are insensitive to the budget wasted on the X-Haul with unnecessarily large capacity.

$E^3$ metric is unable to point out the solution with the best SE, nor the one with the best EE or CE. However, when impacts on different aspects need to be jointly considered, $E^3$ metric provides a feasible way to evaluate the comprehensive gain of the techniques, which contributes to a proper balance among throughput, energy consumption, and expenditure required.

\section{Applications of $E^3$ Metric}

In this section, the impacts of X-Haul and edge cache are represented. The numerical results of different performance metrics are compared as the proof of the rationality of $E^3$ metric. Potential ways to optimize $E^3$ are included as well.

\subsection{$E^3$-Represented Impact of X-Haul}

X-Haul significantly affects the performances of C-RANs, H-CRANs, and F-RANs. Since the maximum capacity of the BS is constrained by the capacity of its X-Haul, X-Haul with larger capacity is preferable while optimizing the throughput of the 5G system. On the other hand, larger capacity also leads to greater expense, which is harmful to the affordability of the 5G system. According to previous researches, cost on X-Haul accounts for 20-40\% of the total operational expenditures when most of the access points (APs) are still money-consuming macro base stations (MBSs) \cite{Backhul_In_HetNet}. Such proportion is liable to increase for 5G systems since most APs are supposed to be much smaller and cheaper to use compared with MBSs, which makes cost on X-Haul a big concern.

Besides, a large portion of the budget is consumed during X-Haul deployment. Considering the flexible deployment, in certain cases it is difficult for LPNs to have access to wired X-Haul, and accordingly the cost on wired X-Haul installation will be too high to afford. In such cases, wireless X-Haul can make a proper alternative, for it requires no extra budget to purchase cables, and the installation is relatively easy to handle, which leads to a lower cost on deployment compared to those wired ones.

X-Haul strategies also affect the energy consumptions of the system. Compared with wired X-Haul, wireless X-Haul consumes more energy, and situations will be even worse when the transmission via wireless X-Haul suffers serious interference.

As it can be observed from the numerical results shown in Fig. \ref{Xhaul}, both SE and EE metrics are insensitive to the overuse of X-Haul capacity resources. Given the limited covering area, there are not as many traffic demands for a LPN as it is for a MBS. Besides, the daily maximum traffic demands can be 2-10 times higher than the daily minimums \cite{EARTH_Report}. That means X-Haul with larger capacity has less chance to get fully used, and excessively pursuing X-Haul with large capacity will lead to a waste of investment. Such results imply that it is preferable to make full use of X-Haul with capacity just passable, for it satisfies the traffic demands to an acceptable level with less expenditure. Actually, some research works focusing on the X-Haul constrained scenarios have already been launched \cite{Fronthaul_Constrained}.

$E^3$ metric also highlights the importance of easy installation. Considering the huge number of deployment, X-Haul solutions for 5G systems should be more cost effective and easy to install than they used to. While both SE and EE are insensitive to the variation of expenditure on X-Haul deployment, $E^3$ can benefit from easy installation as the reduction of expenditure guarantees a lower cost coefficient. As shown in Fig. \ref{Xhaul}, according to $E^3$ performance, the cost-saving feature makes wireless X-Haul an alternative worth considering. Such feature is especially valuable when wired X-Haul brings enormous pressure on the budget of the 5G system. There have already been some research works focusing on exploring the potentiality of wireless X-Haul, and encouraging progresses have been made \cite{Benefit_of_Wireless_Backhaul}.
\subsection{$E^3$-Represented Impact of Edge Cache}
Cache is widely used to reduce the duplicate transmissions of a few popular contents with large sizes. Cache resources used to be mostly deployed within mobile content delivery networks. Such centralized manner makes it easy to improve the visibility of both content dynamics and user characteristics. However, the traffic data over X-Haul still surges a lot of redundant information, which worsens the X-Haul constraints. Besides, the reduction of user latency is also limited, as a certain portion of the delay is caused by the X-Haul. To further optimize the performance of the system, a substantial amount of cache resources begin to be deployed in a distributed manner. By moving application processing resources toward the network edge closer to mobile users, it simultaneously eases the pressure on X-Haul and reduces the latency of users.

Cache helps to improve the capability of the system as well as the quality of services, but it puts challenging requirements on the placement and utilization cost as well. Considering the explosive growth of mobile traffics, the demands for cache resources will keep increasing, which makes the cost on cache become ever more important.

\begin{figure}[t]
\centering
\vspace*{0pt}
\includegraphics[scale=0.65]{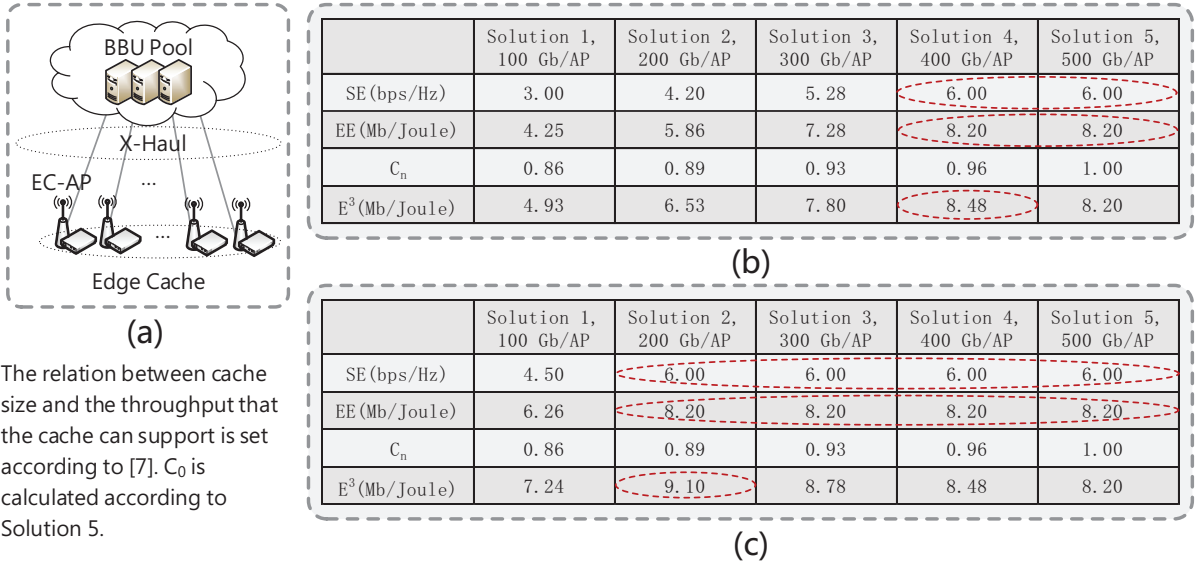}
\caption{Different performance metric comparisons: the impact of edge cache. (a) An example of F-RAN based 5G systems. (b) Numerical results for a referential caching strategy. (c) Numerical results for an advanced caching strategy.}
\label{Cache}
\end{figure}

For different caching strategies, Fig. \ref{Cache} shows a comparison among SE, EE, and $E^3$ metrics. The size of cache equipped to each AP varies with solutions. Besides, two different caching strategies are involved, and the one used in Fig. \ref{Cache} (c) is assumed to have better performance. The capacity of X-Haul is set to $c_1$, which is less than enough to fully meet the traffic demands without the help of edge cache. For simplicity, the energy consumed by cache is ignored.

As shown in Fig. \ref{Cache}, all three metrics highlight the use of edge caching technique, as it improves the performance when the capacity of X-Haul is the handicap of system throughput. However, while $E^3$ can point out the overuse of edge cache resources, traditional SE and EE metrics are insensitive to such problem. Besides, as it can be observed from the comparison between Fig. \ref{Cache} (b) and Fig. \ref{Cache} (c), $E^3$ metric also highlights the importance of advanced caching strategy, as the performance will be better if the cache resources are used more efficiently. Another fact worth noticing is that only a small amount of popular contents are accessed by a large portion of requests, while others remain unpopular. Considering the scale of contents is expanding at an explosive speed, it is not possible to cache all contents at BSs. As a result, compared with expanding the cache size, more attention is required by the improvement of caching strategies to feasibly optimize the $E^3$ performance. Actually, there have already been some research works focusing on this topic, and some promising achievements have been proposed \cite{Introduction_to_Cache}.
\subsection{The Joint $E^3$-Represented Impact of X-Haul and Edge Cache}
Both X-Haul and edge cache are essential components of F-RAN based 5G system, and the corresponding strategies also significantly impact the performance of each other. On one hand, edge cache techniques help to relieve the pressure on X-Haul. On the other hand, with more transmission supported by X-Haul, the demands for edge cache resources will decrease, and the edge cache can be used more efficiently. Putting these features together, the F-RAN based 5G system can meet the traffic demands with less budget.

\begin{figure}[t]
\centering
\vspace*{0pt}
\includegraphics[scale=0.65]{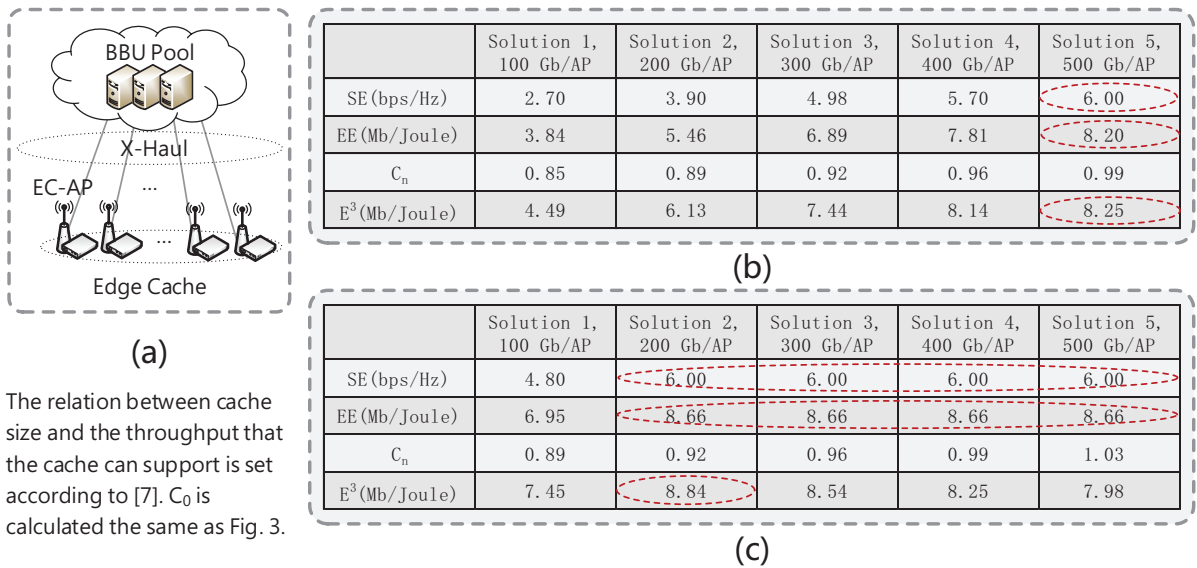}
\caption{Different performance metric comparisons: the joint impact of X-Haul and edge cache. (a) An example of F-RAN based 5G systems. (b) Numerical results for solutions with X-Haul capacity = $c_2$ (c) Numerical results for solutions with X-Haul capacity = $c_3$.}
\label{Joint}
\end{figure}

For different combinations of X-Haul and edge caching strategies, Fig. \ref{Joint} shows a comparison among SE, EE, and $E^3$ metrics. The capacity of X-Haul are set to $c_2$ and $c_3$ ($c_2<c_1<c_3$), respectively. Both $c_2$ and $c_3$ are less than enough to fully meet the traffic demands without the help of edge cache. Caching strategies used in Fig. \ref{Joint} (b) and Fig. \ref{Joint} (c) are assumed to be the same as the one used in Fig. \ref{Cache} (b).

As it can be observed from the comparisons among Fig. \ref{Cache} (b), Fig. \ref{Joint} (b), and Fig. \ref{Joint} (c), according to EE, a joint optimization is not an issue worth considering, since the optimal performance remains the same as long as there are enough edge cache resources to guarantee the maximum throughput. Nevertheless, different conclusion is made according to $E^3$. The numerical results imply that the optimal $E^3$ performance and the optimal cache size will be affected by the change in X-Haul. Consequently, a joint consideration on both X-Haul and edge cache strategies deserves attention while optimizing $E^3$.

\section{Challenges and Open Issues}

$E^3$ metric provides a feasible and comprehensive way to evaluate the performances of F-RAN based 5G systems when new elements like X-Haul and edge cache are taken into consideration. However, as shown in Fig. \ref{Challenge}, there are still challenges ahead, including the model of cost, $E^3$ optimization based radio resource allocation, and $E^3$ optimization for IoTs.

{\color{red}
\begin{figure}[t]
\centering
\vspace*{0pt}
\includegraphics[scale=0.9]{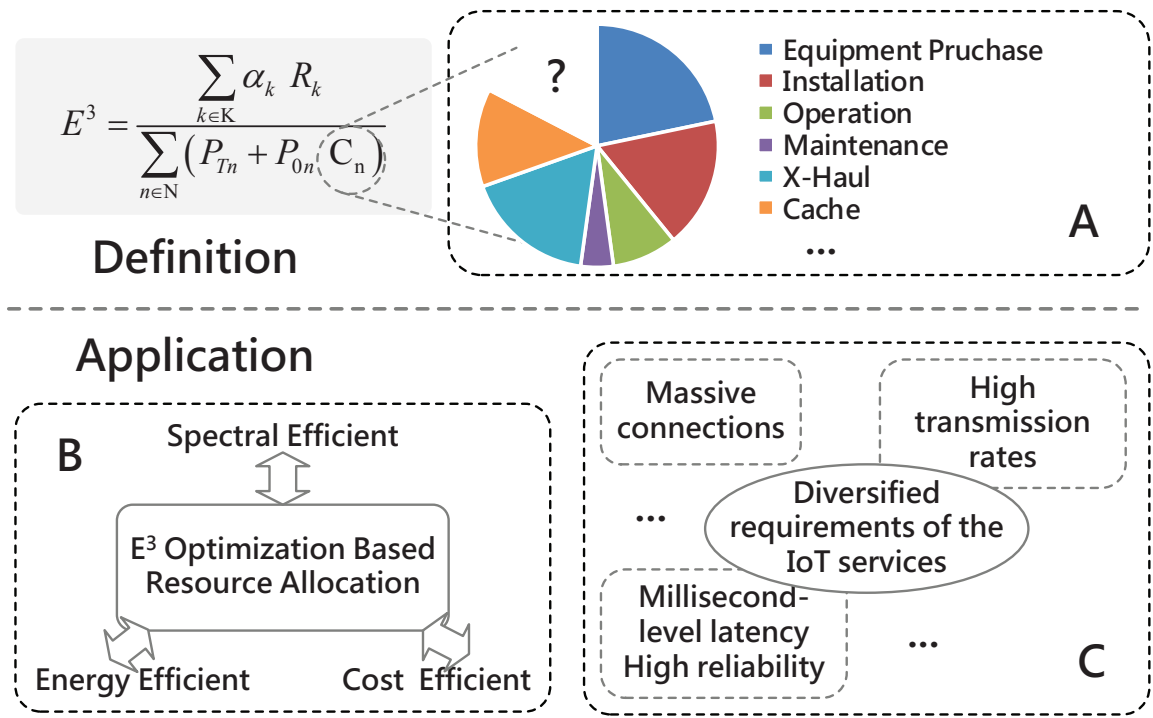}
\caption{Potential challenges and open issues for $E^3$ metric}
\label{Challenge}
\end{figure}
}

\subsection{Cost Modelling}

$E^3$ metric aims to provide a feasible way to fully consider the impacts on different aspects. As an enhancement to traditional EE metric, the introduction of cost coefficient is essential for considering the cost impact. Consequently, the calculation of cost coefficient is crucial for $E^3$ metric, which makes it a key issue to obtain the precise characterization of the cost.

However, according to the authors' knowledge, there has not been a method of modeling the cost that is applicable to $E^3$ metric. Besides the sophisticated constitution of the cost, the impact of the existing equipment from traditional networks also needs to be considered, which makes the problem even more complicated. New techniques like network function virtualization and cloudization also increase the difficulty of solving the problem. Considering the importance and the complexity, modeling the cost is an urgent topic for future research works.

\subsection{$E^3$ Optimization Based Radio Resource Allocation}

According to $E^3$ metric, increasing throughput or reducing energy consumption alone is not enough to make a radio resource allocation strategy with better performance. Specifically, a proper balance between the capability of the system and the corresponding overheads is preferable, which requires effective use of various resources of the 5G system, including spectral resources, energy, and budget.

Without a feasible way to take the impacts on throughput, greenness, and affordability together into consideration, traditional techniques used to be designed mainly focusing on one or two aspects. As a result, few existing techniques can guarantee the comprehensive improvement of performance. Since high SE, high EE and high CE are regarded as the three essential features of 5G systems, current radio resource allocation strategies have to be evolved in an $E^3$ optimization manner to appropriately meet the demands in the future.

\subsection{$E^3$ Optimization for IoTs}

Recently, IoT has become the main driver in the future development of mobile communications, which extending the scope of mobile communication services from interpersonal communications to smart interconnections between things and things, and between people and things. With the emergence of IoTs, mobile communication technologies now can penetrate into broader industries and fields.

IoT brings about infinite vitality to mobile communications, as well as great challenges. As shown in Fig. \ref{Challenge}, the demands of diversified IoT services are significantly different, and 5G systems have to meet such demands in a flexible and sustainable manner, which takes great challenges to the 5G system. Accordingly, it is a challenging concern to optimize the $E^3$ performance of IoTs, which is urgent to be exploited for the successful rollout of IoTs in 5G sytems.

\section{Conclusion}

To characterize the scalability, flexibility, and interoperability, an advanced $E^3$ metric is proposed in this article to evaluate the impacts of X-Haul and edge cache in the F-RAN based 5G systems. With the traditional EE and the cost taken into account, the proposed $E^3$ metric provides a feasible way to show comprehensive gains when different kinds of advanced technologies are used. Based on the numerical results, ways to optimize $E^3$ performance of 5G systems are included in this article. It is concluded that $E^3$ metric serves as a proper choice when the impacts on throughput, greenness, and affordability all require consideration. However, being a new proposed performance metric, there are still a number of problems urgent to be solved in the future, and special attention is required by the key issues including the model of cost, $E^3$ optimization based radio resource allocation, and $E^3$ optimization for IoTs.

\end{document}